\documentclass[final,3p,times]{elsarticle}
\usepackage{amssymb} 					%amssymb package provides various useful mathematical symbols
\usepackage{amsmath} 					%amsmath package provides various useful math options
\usepackage[latin1]{inputenc} %mutated vowels (german)
\usepackage{url}							%url for websites
\usepackage{rotating}
\usepackage{lscape}
\usepackage{float}        		%floats with also H not only h,t,b,p
\journal{Statistical Mechanics and its Applications}
\begin{document}
\begin{frontmatter}
\title{From invasion percolation to flow in rock fracture networks}
\author[ETH]{Salomon J. Wettstein}
\author[ETH]{Falk K. Wittel}
\author[ETH]{Nuno A. M. Ara\'{u}jo}
\author[NAGRA]{Bill Lanyon}
\author[ETH]{Hans J. Herrmann}
\address[ETH]{Computational Physics for Engineering Materials, IfB, ETH Zurich, Schafmattstr. 6, CH-8093 Zurich, Switzerland}
\address[NAGRA]{Fracture Systems Ltd., St. Ives, Cornwall, United Kingdom}
\begin{abstract}
  The main purpose of this work is to simulate two-phase flow in the form of immiscible displacement
	through anisotropic, three-dimensional (3D) discrete fracture networks (DFN). The considered DFNs are
	artificially generated, based on a general distribution function or are conditioned on measured data
	from deep geological investigations. We introduce several modifications to the invasion percolation (MIP)
	to incorporate fracture inclinations, intersection lines, as well as the hydraulic path length inside the fractures.
	Additionally a trapping algorithm is implemented that forbids any advance of the invading fluid into a region,
	where the defending fluid is completely encircled by the invader and has no escape route.
	We study invasion, saturation, and flow through artificial fracture networks, with varying anisotropy and
	size and finally compare our findings to well studied, conditioned fracture networks.
\end{abstract}
\end{frontmatter}
%~~~~~~~~~~~~~~~~~~~~~~~~~~~~~~~~~~~~~~~~~~~~~~~~~~~~~~~~~~~~~~~~~~~~~~~~~~~~~~~~~~~~~~~~~~~~~~%

%~~~~~~~~~~~~~~~~~~~~~~~~~~~~~~~~~~~~~~~~~~~~~~~~~~~~~~~~~~~~~~~~~~~~~~~~~~~~~~~~~~~~~~~~~~~~~~%
% sections
%~~~~~~~~~~~~~~~~~~~~~~~~~~~~~~~~~~~~~~~~~~~~~~~~~~~~~~~~~~~~~~~~~~~~~~~~~~~~~~~~~~~~~~~~~~~~~~%

%%%%%%%%%%%%%%%%%%%%%%%%%%%%%%%%%%%%%%%%%%%%%%%%%%%%%%%%%%%%%%%%%%%%%%%%%%%%%%%%%%%%%%%%%%%%%%%%
\section{Introduction}\label{sec: sec_introduction}
%%%%%%%%%%%%%%%%%%%%%%%%%%%%%%%%%%%%%%%%%%%%%%%%%%%%%%%%%%%%%%%%%%%%%%%%%%%%%%%%%%%%%%%%%%%%%%%%

Fracture networks are abundant in natural geological formations. Characterizing and investigating
them is among the most challenging problems in geosciences, ranging from ground-water hydrology,
subsurface oil, isolation of nuclear and hazardous wastes or captured CO$_2$, to geotechnical problems
involved in tunneling and cavern excavations. The identification and assessment of fractures
as capillary barriers, the understanding of flow and transport in fracture networks, as well as the development
of mathematical models is crucial for predicting the hydraulic behavior of fractured geological systems.
Due to the geometrical complexity of the natural system, analytical solutions are not available,
but also it is not feasible to describe the systems in detail \cite{Neuman2005,Suess2004}.
To be able to study natural systems, their structure, as well as the occurring processes, they
have to be represented by conceptual models. Numerical solutions have to be applied which involve
spatial and temporal discretization of the problem and usage of efficient and stable algorithms.

One approach to modeling flow in fractured rock is to represent the fracture network within the rock as
a three-dimensional discrete fracture network (DFN). The DFNs are either given on the basis of
core samples and measurements on exposures (tunnel walls or outcrops), or they emerge from planar, irregular,
$n$-polygonal fractures that are oriented around the mean direction, which may follow for example
a Fisher distribution \cite{Fisher1953}. The latter are considered to be a solid representation
of the ones experimentally measured \cite{Dershowitz1988,Priest1993,Adler1999,Kemeny2003}. One
possible approach for modeling the displacement of immiscible fluids like gas, oil, or water inside
DFN models, is the so-called \textit{invasion percolation} (IP).
IP is a dynamic growth process that generates phase structures through a set of rules, which
embody the physics of immiscible displacement of fluids within a random field, such as a network of
pores or fractures. IP was proposed by Lenormand and Bories \cite{Lenormand1980} and Chandler \textit{et al.}
\cite{Chandler1982} in the early 1980s, as well as Wilkinson and Willemsen who coined the name IP
\cite{Wilkinson1983}. They emphasized that IP is a modified form of ordinary percolation (OP)
\cite{Broadbent1957}, only with a well-defined sequence of invasion events. In IP it is assumed that capillary
forces dominate the viscous ones. In other words, both gravity and viscous effects are neglected.
Nevertheless, IP is a valid approximation for the flow physics and well suited to describe the
slow immiscible displacement of two fluids in porous media or fracture networks. In this work,
physical modifications to the model are proposed to study the gas-water displacement process in
the DFNs. IP has been \textit{modified} (MIP) in many ways in the past to incorporate more geometrical and
physical details of the network like force fields such as gravity \cite{Meakin1992,Glass1996,Ioannidis1996,Oliveira2010},
centrifugal forces \cite{Holt2003}, viscous forces \cite{Xu1998}, as well as capillary smoothing
mechanisms in porous media \cite{Glass1996,Blunt1995} and inside single fractures \cite{Glass1998}.
In this work, we have modified the IP model to a more realistic description of the invading progress.
To decide if the invader will penetrate the fracture adjacent to the invading front, not only the
action of a fracture as a capillary barrier, but also the contact profile to this
fracture and its inclination with respect to the flow direction, as well as the hydraulic path length
from entry to exit will be considered. As the invasion proceeds, part of the invasion percolation
cluster boundary might completely encircle a region of the fracture network. This phenomenon, named trapping,
calls for a sophisticated IP that forbids any advance into this area. The IP or MIP routine is
appropriately called trapping IP \cite{Wilkinson1983} or trapping MIP. To study leakage rates and
pressure fields above percolation, after pruning dead ends (backbone of the percolating cluster),
the cubic law for the remaining fracture network is solved in the form of a linear hydraulic
system. The global system matrix is composed of invaded fractures as elements and the
connections as nodes using a simple channel model following Cacas \textit{et al.} \cite{Cacas1990}.

The paper is organized as follows:
Sec.~\ref{sec: sec_fracture_networks} provides information on 3D fracture networks in general and
describes the generation of discrete fracture network (DFN) models based on either a general distribution
function or on previously measured distributions. The physical modifications of the invasion percolation
(MIP) procedure including the trapping phenomenon are introduced in Sec.~\ref{sec: sec_physical_modifications},
followed by the explanation of the typical invasion scenarios in Sec.~\ref{sec: sec_typical_inv_scenarios}.
Sec.~\ref{sec: sec_permeability_breakthrough} is devoted to the physics of flow in single fractures and
subsequently in fracture networks. In Sec.~\ref{sec: sec_isotropic_anisotropic} the results of simulations
in artificially generated fracture networks with Fisher distributed fracture orientations are presented and
complemented by a size dependence study of the relevant properties in Sec.~\ref{sec: sec_finite_size_dependence}. Finally,
results are compared with those from a model conditioned on data from geological investigations \cite{Mazurek1998}
in Sec.~\ref{sec: sec_comparison_wellenberg}. The paper closes with conclusions and an outlook in
Sec.~\ref{sec: sec_conclusions}.

%%%%%%%%%%%%%%%%%%%%%%%%%%%%%%%%%%%%%%%%%%%%%%%%%%%%%%%%%%%%%%%%%%%%%%%%%%%%%%%%%%%%%%%%%%%%%%%%
\section{Fracture networks}
	\label{sec: sec_fracture_networks}
%%%%%%%%%%%%%%%%%%%%%%%%%%%%%%%%%%%%%%%%%%%%%%%%%%%%%%%%%%%%%%%%%%%%%%%%%%%%%%%%%%%%%%%%%%%%%%%%

Natural geological systems are highly complex in terms of the geological structure as well
as the physical processes occurring within them, making analytical solutions unfeasible.
Hence, a simplifying transformation of the natural system to a numerical model is required \cite{Suess2004}.
The natural fracture system has to be transformed to a discrete fracture network (DFN)
that is composed of individual, intersecting fractures with similar hydraulic characteristics
as the real system. To be able to make systematic studies, that are comparable to the state of the
art knowledge, the artificial DFNs are constructed in a similar manner to those of Refs.~\cite{Huseby1997,Khamforoush2007}.

\begin{figure}[H]
	\centering
	\includegraphics[width=0.41\textwidth]{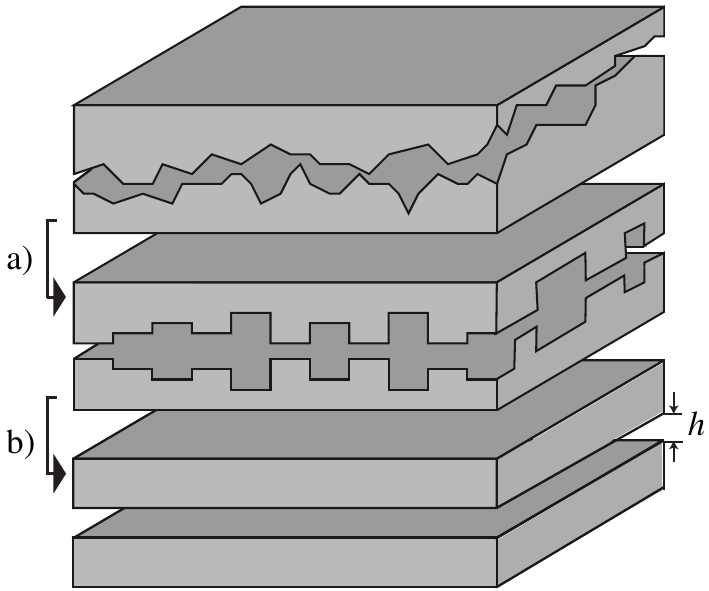}
  \caption[Parallel-plate concept]
	{Parallel-plate concept:
		 a) From natural single fracture to local parallel plates.
		 b) From local parallel to one single plate with mean aperture $h$ (distance between the plates)
		 	  following \citet{Suess2004}.}
	\label{fig: fig_parallel_plate_concept}
\end{figure}

The hydraulic properties of fracture network systems are mostly determined by the permeability of the
fractures and matrix. In our conceptual model, the permeability is exclusively determined by fractures.
The matrix between the fractures is not considered, since its permeability compared to
the ones of the fractures is assumed to be extremely low (see \cite{Lanyon2009}). The hydraulic properties of the DFN
are typically characterized by a fracture size, fracture permeability, fracture orientation, and fracture
density distribution. Individual fractures can be seen as a two-dimensional porous medium with flow
inside the fracture plane. Variations of pressure or velocity across the height $h$, called the aperture
of the fracture, can be averaged out. The aperture is much smaller than the other two dimensions of the
fracture and has an essential influence on the flow and transport processes in fracture networks. In the
DFN model, the aperture is described by the parallel-plate concept (Fig.~\ref{fig: fig_parallel_plate_concept}),
that is the simplest model of flow through a single fracture with the possibility for
exact solutions of the hydraulic conductivity.

%----------------------------------------------------------------------------------------------%
\subsection{Artificial fracture network models}
	\label{subsec: subsec_AFN_DFN}
%----------------------------------------------------------------------------------------------%

The artificial DFN is composed of $N_P$ planar, regular or irregular, convex polygons with
$N_V$ vertices. Each fracture is approximated by an individual polygon $i$ (with $i=1\ldots N_P$)
inscribed in a disk of radius $R_i$. This approximation comes from the parallel plate model, assuming
that rigid walls are smooth and plain. The radii $R_i$ of the circumscribed circles are randomly chosen
in the interval $\left[R_\text{min},R_\text{max}\right]$, and the number of vertices $N_V$
is picked in the interval $\left[3,N_\text{max}\right]$, both from a uniform distribution. The vertices are distributed
on the perimeter of the circle with uniformly distributed angles $\alpha_1,\alpha_2, \ldots ,\alpha_{N_V}$,
all in the interval $\left[0,2\pi\right]$. A simplification can be made in terms of choosing regular,
equal sized (monodispersed) polygons on circles with fixed radii $R_i=R_\text{max}$, as well as an
identical number of vertices $N_V$ and subsequently equal angles $\alpha_j=2\pi/N_V$.
The aperture is in this case just the ratio of the pore volume $V_{\text{poly},i}$ over the area $A_{\text{poly},i}$.
The pore volume $V_{\text{poly},i}$ is calculated using $V_{\text{poly},i} = A_{\text{poly},i}^{3/2} z_i V_{\text{char}}$,
with the uniformly distributed random variable $z_i\in\left[0,1\right]$ and the characteristic
pore volume $V_{\text{char}}$, which is a given constant depending on the fracture system under consideration.
Detailed investigations of fractures suggest that aperture varies within each fracture (e.g. \cite{Abelin1994}).
However, within the current model we assume the fracture aperture to be uniform over each polygon. 
Hydrogeological investigations, where transmissivity or equivalently hydraulic aperture is assumed to be uniform within fractures
and is determined by pumping tests, typically show a long-tailed aperture distribution for all fractures, often represented as a log-normal
or Pareto distribution \cite{Gustafson2005}.

Once all polygonal fractures are generated, their centers are placed in the simulation box of size $L^3$,
using a uniform distribution before orientations are adjusted. The homogenization scale $L$ is chosen such
that it significantly exceeds the size of the largest possible fracture $k$ by $L\gg d_k$, with $d_k=d_\text{max}$
the diameter of the circumscribed circle of fracture $k$. The fracture orientation is represented by the normal vector of the
fracture plane. The distribution of a random set of normal vectors can be described by means of a probability density
function $f(\theta,\phi)$, expressed in standard spherical coordinates $\theta$ and $\phi$. The simplest probability
density function would be the uniform distribution of the normal vectors on the unit sphere. However, in order to study anisotropic,
three-dimensional (3D) DFN, which are closer to reality, the \textit{Fisher distribution} can be used \cite{Khamforoush2007,Khamforoush2008}.
This distribution is the analogue of the Gaussian distribution on the sphere \cite{Fisher1953}
and can be deduced in polar coordinates $\theta\in\left[0,\pi\right]$ and $\phi\in\left[0,2\pi\right]$
as follows:
\begin{equation}
	f(\theta,\phi,\kappa) = \frac{\kappa}{4\pi\sinh\kappa} \sin\theta
	\exp{\left(\kappa\left[\cos\theta_0\cos\theta+\sin\theta_0\cos(\phi-\phi_0)\right]\right)} \ ,
	\label{eq: eq_fisher_distr1}
\end{equation}
with $\kappa\ge 0$ denoting the concentration or \textit{dispersion parameter}. For the particular case where the initial
coordinates $\theta_0$ and $\phi_0$ are equal to zero, the Fisher distribution reduces to
\begin{equation}
	f(\theta,\kappa) = \frac{\kappa}{4\pi\sinh\kappa} \sin\theta\exp{\left(\kappa\cos\theta\right)} \ .
	\label{eq: eq_fisher_distr2}
\end{equation}
This distribution is rotationally symmetric around the initial mean direction, which coincides
with the $z$-axis. The greater the value of $\kappa$, the higher the concentration of the distribution
around this axis. The distribution is unimodal for $\kappa>0$ and is uniform on the sphere for $\kappa=0$.

Examples of generated artificial 3D DFNs with different $\kappa$ values are given in Figs.~\ref{fig: fig_AFN_DFN_model}
a) and b). Note that only when fractures are highly connected, the system behaves like a continuous medium.
Since the interest lies in the movement of a fluid phase in the network, the fracture density $\rho_{\text{fr}}$
of a DFN, being the number of fractures per volume, has to be high enough to build a spanning cluster over the entire
domain. The connectivity of the DFN can be resembled by the dimensionless density or the so-called concentration $\rho^{'}$.
Huseby \textit{et al.} \cite{Huseby1997} have shown, that the concentration $\rho^{'}$ equals the average number of
intersections per fracture, $\rho^{'}=\left\langle\overline{N_\text{int}}\right\rangle$. The concentration is
thereby a decreasing function of $\kappa$ \cite{Khamforoush2007}. Hence, a much smaller average number of
intersections per fracture is expected for the highly anisotropic DFN (see Fig.~\ref{fig: fig_AFN_DFN_model} b))
than for the nearly isotropic one (see Fig.~\ref{fig: fig_AFN_DFN_model} a)). Furthermore, Khamforoush \textit{et al.}
\cite{Khamforoush2008} noted that the asymptotic values (for the infinite system $L\rightarrow\infty$)
of percolation thresholds (of OP) for anisotropic DFNs ($p_c\equiv \rho^{'}_c$) are in
\begin{equation}
	2.1 \leq \rho^{'}_{c,\infty,x,A} ,~ \rho^{'}_{c,\infty,y,A} \leq 2.3 \qquad \text{and} \qquad
	2.3 \leq \rho^{'}_{c,\infty,z,A} \leq 2.44 \ .
	\label{eq: eq_rho_crit_2}
\end{equation}
$\rho^{'}_{c,\infty}$ denote critical concentrations in the range of $0\leq\kappa\leq 50$ with
the mean direction of the Fisher distribution chosen to be the $z$-axis.

%----------------------------------------------------------------------------------------------%
\subsection{Realistic fracture network models}
	\label{subsec: subsec_WLB_DFN}
%----------------------------------------------------------------------------------------------%

In addition to the artificial networks described in the previous section it was thought useful to
consider more realistic models conditioned on field data. Within the study, network models of the
Excavation Damage Zone around tunnels in Opalinus Clay (not described here) and a large-scale model
of the fracture system within a fractured marl were considered. The large-scale model is a simplified
version of that described in Mazurek \textit{et al.} \cite{Mazurek1998} and Nagra \cite{Nagra1997}.
It was selected as a model likely to show a well-connected 3D fracture system suitable for comparison
with the analytic models. The model is conditioned on field data acquired from deep boreholes in the
Valanginian Marl at Wellenberg in Central Switzerland during Nagra's investigations there in the early
1990s \cite{Nagra1997}. The model includes fracture ``sets'' corresponding to the inventory of water
conducting features (WCFs) identified in the boreholes. The WCFs included cataclastic faults zones over
a wide range of length scales and smaller structures including thin discrete shear zones, joints and
isolated limestone beds. The full model also included a small number of larger-scale limestone banks that
had been identified but these were not in the stochastic model used here. The length distribution of the
cataclastic zones was based on the observed thickness and roughly followed a power-law form with exponent
between $2.3$ and $2.6$. Feature orientation was based on observations from core and showed a strong
relationship to the folding axis of the rock but includes discordant sets resulting in a well connected
network. Hydraulic aperture distributions were  calculated from the observed transmissivity measured by
pumping tests and fluid logging analyses.

\begin{figure}[H]
	\includegraphics[width=1.0\textwidth]{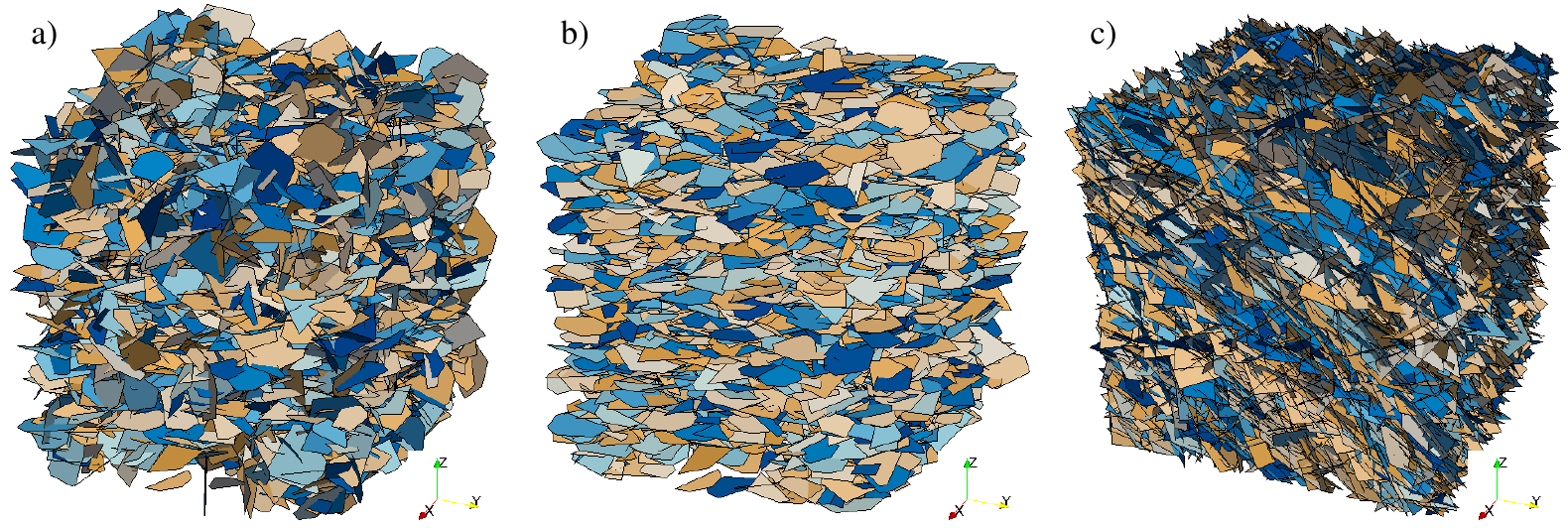}
	\caption[MIP on an isotropic artificial 3D DFN - breakthrough]
		{a) Nearly isotropic ($\kappa=1$) artificial 3D DFN with system size $L=10\ d_\text{max}$,
		 where $d_\mathrm{max}$ is the diameter of the circumscribed circle of the largest fracture in the DFN,
		 containing $4000$ fractures.
		 b) Highly anisotropic ($\kappa=50$) artificial 3D DFN with the same size and number of fractures.
		 c) Sample of the Wellenberg DFN with system size $L=100$~m containing $12684$ fractures.}
	\label{fig: fig_AFN_DFN_model}
\end{figure}

The Wellenberg DFN model shown in Fig.~\ref{fig: fig_AFN_DFN_model} c) is comparable to the 3D randomly
generated artificial fracture networks and is built similarly in the form of a $L^3$
system with $L=100$~m. Large natural faults, spanning over hundreds of meters, are thereby tessellated
into smaller fractures, represented by polygons with side length
of $R_i=10$~m, using again the parallel plate model. The concentration in terms of the average number
of intersections per fracture is $\rho^{'}\approx 15$ and thus much higher than for the artificially
generated DFN. This leads to a higher connectivity of the DFN and hence only a small amount of water
is assumed to be trapped in the invasion percolation simulation afterwards.

%%%%%%%%%%%%%%%%%%%%%%%%%%%%%%%%%%%%%%%%%%%%%%%%%%%%%%%%%%%%%%%%%%%%%%%%%%%%%%%%%%%%%%%%%%%%%%%%
\section{Physical modifications for IP}
	\label{sec: sec_physical_modifications}
%%%%%%%%%%%%%%%%%%%%%%%%%%%%%%%%%%%%%%%%%%%%%%%%%%%%%%%%%%%%%%%%%%%%%%%%%%%%%%%%%%%%%%%%%%%%%%%%

IP in its simplest form fully invades an entire fracture, once the aperture threshold is reached
and at least one intersecting fracture is filled. For fluid invasion, previous experiments have
shown that the detailed behavior at fracture intersections are essential \cite{Sarkar2004}. We use a
modification to explicitly represent intersections and to implement rules at fracture intersections
to mimic capillary barrier type behavior for the invading gas in form of an adjusted entry aperture
$h^{\text{E}}_\text{adj}$. Given that some fractures of the fracture network models can be much
bigger than others and that maybe only a small portion of them will be invaded, an additional
modification considering the path length inside the fracture for the hydraulic aperture $h^{\text{H}}$
of a fracture is suggested.

\begin{figure}[H]
	\centering
	\includegraphics[width=0.5\textwidth]{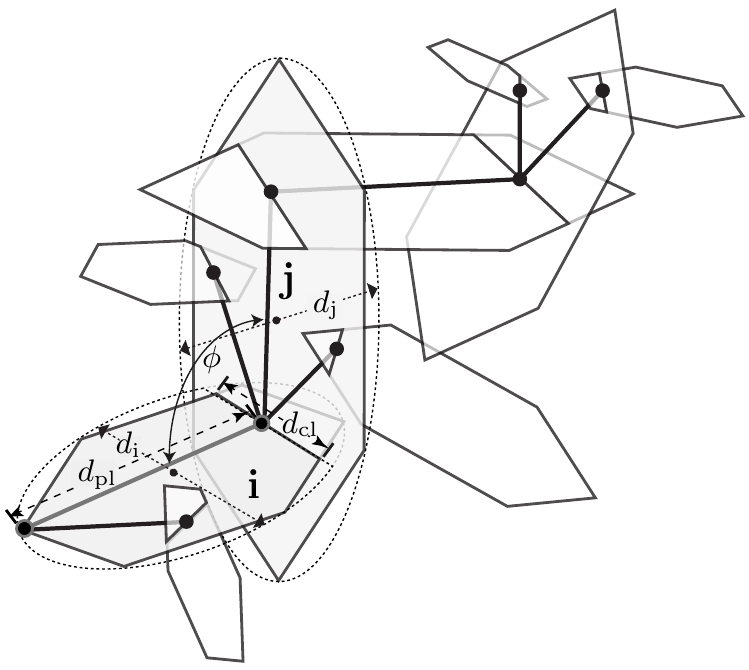}
	\caption[Comparison IP to MIP]
		{Modified invasion percolation (MIP) with invaded fracture $i$ at the invading front connected
		 to the non-invaded fracture $j$ (candidate for invasion). $d_{\text{pl}}$ is the path length
		 in fracture $i$ from prior (invaded) entry node to the exit node to fracture $j$ and $d_{i}$ is
		 the diameter of the circumscribed circle of fracture $i$. $d_{\text{cl}}$ is the length of the
		 line of intersection between fracture $i$ and $j$, $d_{j}$ is the diameter of the circumscribed
		 circle of fracture $j$ and $\phi$ is the contact angle between the two fractures.}
	\label{fig: fig_MIP_adjustments}
\end{figure}

Consider an invaded fracture $i$ from the invasion front, connected to a non-invaded fracture $j$,
that is a feasible candidate for invasion. The resistance of fracture $j$ is given by its
entry aperture $h^{\text{E}}_\text{j}$, which has to be at least as high as the current threshold
value to allow for invasion. In the following, the aperture value of fracture $j$ will be adjusted
with respect to the effects mentioned above:
\begin{itemize}
	\item The length of the \textit{line of intersection} $d_{\text{cl}}$ between fracture $i$ and $j$, which results
				from the connectivity calculation, is divided by the diameter of the circumscribed circle $d_{j}$ taken as
				the maximal width of fracture $j$ (see Fig.~\ref{fig: fig_MIP_adjustments}). Hence the discriminating factor
				concerning the line of intersection is
				\begin{equation}
					\text{n}_{\text{C}} = \displaystyle\frac{d_{\text{cl}}}{d_{j}} \ .
					\label{eq: eq_intersection_factor}
				\end{equation}
				The corresponding factor for the invasion of entry fractures at the entry zone of the fluid is $n_{\text{C}}=1$.
	\item To consider the \textit{inclination} of the fractures with respect to the flow direction, the aperture is corrected
				with a discriminating factor $\cos{(\phi)}$ (where $0 \leq\phi\leq\pi/2$) as proposed by Sarkar \textit{et al.}
				\cite{Sarkar2004}. The angle $\phi$ is either the contact angle between fracture $i$ and $j$ at the invading
				front (see Fig.~\ref{fig: fig_MIP_adjustments}) or the angle between the global inflow direction and the inclination
				of the considered entry fracture at the entry zone. This results for both cases in the discriminating factor of
				\begin{equation}
					\text{n}_{\text{I}} = \cos{\left(C_{\text{I}} \phi\right)} \ ,
					\label{eq: eq_inclination_factor}
				\end{equation}
				with the adjustment parameter $C_{\text{I}}$ between $0$ and $1$, to adjust the strength of the inclination
				effect. Since $\cos{\left(\phi\right)}=0$ for $\phi=\pi/2$, the discriminating factor vanishes for fractures
				perpendicular to each other and flow between them would be prohibited. This is anticipated by setting $C_{\text{I}}<1$.
				In the other extreme, no inclination adjustment is obtained by setting $C_{\text{I}}=0$.
	\item The \textit{path length} adjustment is done using the path length discriminating factor
				\begin{equation}
					\text{n}_{\text{P}} = \displaystyle\frac{d_{i}}{d_{\text{pl}}} \ ,
					\label{eq: eq_pathlength_factor}
				\end{equation}
				where $d_{\text{pl}}$ is the path length in fracture $i$ from prior (invaded) entry node to the exit node to fracture $j$.
				$d_{i}$ is the diameter of the circumscribed circle of fracture $i$ being the maximal possible path length
				(see Fig.~\ref{fig: fig_MIP_adjustments}).
\end{itemize}
Finally the \textit{adjusted entry} and \textit{adjusted hydraulic aperture} of fracture $j$ for invasion from
fracture $i$ is
\begin{equation}
	h^{\text{E}}_\text{adj,j} = \text{n}_{\text{C}} \text{n}_{\text{I}} h^{\text{E}}_\text{j}
		\quad \text{and} \quad
	h^{\text{H}}_\text{adj,j} = \text{n}_{\text{P}} h^{\text{H}}_\text{j} \ .
	\label{eq: eq_adjusted_eaperture_and_haperture}
\end{equation}
The ordinary IP is retrieved by the special case where $\text{n}_{\text{P}}=\text{n}_{\text{C}}=\text{n}_{\text{I}}=1$.
The resulting total aperture is calculated using the rule for computing the equivalent aperture
for fractures \textit{connected in series} \cite{Wilson1974}
\begin{equation}
	h^{\text{T}}_\text{adj,j} = \displaystyle\frac{1}{(h^{\text{E}}_\text{adj,j})^{-1}+(h^{\text{H}}_\text{adj,j})^{-1}} \ .
	\label{eq: eq_total_aperture}
\end{equation}

%----------------------------------------------------------------------------------------------%
\subsection{Searching for traps}
	\label{subsec: subsec_TMIP_trapping_algorithm}
%----------------------------------------------------------------------------------------------%

As the MIP proceeds, part of the cluster boundary might completely encircle a region of the
fracture network containing the defending fluid (see Fig.~\ref{fig: fig_infinite_cluster_scheme}).
Given that the fluid is nearly incompressible, the entrapped region cannot be invaded.
Hence the MIP version forbidding any advance into such an area is appropriately called
\textit{trapping} MIP (\textit{TMIP}). Searching for traps and removing their perimeters from the
list of potential invasion sites is cumbersome, but as a matter of fact one of the most important
parts of IP simulations \cite{Sahimi1994}. In early attempts, the entire lattice has been scanned
using a \textit{Hoshen-Koplemann} (HK) algorithm \cite{Sahimi1994}, which finds and labels all
connected, water-filled regions that are disconnected from the outlet and scales as $\mathcal{O}(N)$.
Nevertheless, the computation time for the whole trapping algorithm
is costly in computing time and scales as $\mathcal{O}(N^2)$ with the time for each lattice realization.
For this reason Sheppard \textit{et al.} \cite{Sheppard1999} designed a highly efficient algorithm whose execution time
scales as $\mathcal{O}(N \ln N)$. After insuring that trapping can occur, they used several
simultaneous so-called breadth-first searches to update the cluster labeling. This is similar
to the strategy considered here, where we use the burning algorithm, first proposed by
Herrmann \textit{et al.} \cite{Herrmann1984}. To check the possibility of trapping, the last
invaded fractures are included in a list of potentially trapped fractures. For each fracture in
this list, a so-called depth-first search is used to check, for connectivity to the outlet.
Fractures that are designated as trapped are labeled and removed from the invasion list.
The efficiency of this algorithm stems from the fact that the number of traps decreases as a power
law with the trap size \cite{Sahimi1994} and small traps can be detected quickly through local searches.

%%%%%%%%%%%%%%%%%%%%%%%%%%%%%%%%%%%%%%%%%%%%%%%%%%%%%%%%%%%%%%%%%%%%%%%%%%%%%%%%%%%%%%%%%%%%%%%%
\section{Typical invasion scenarios}
	\label{sec: sec_typical_inv_scenarios}
%%%%%%%%%%%%%%%%%%%%%%%%%%%%%%%%%%%%%%%%%%%%%%%%%%%%%%%%%%%%%%%%%%%%%%%%%%%%%%%%%%%%%%%%%%%%%%%%

Before the actual MIP simulation can be started, one has to determine, if the network percolates.
Therefore a depth-first search called \textit{forward percolation} is performed to ensure that the
inlet is connected to the outlet. The inlet is defined by an entry zone and the outlet by an
exit zone with a predefined volume expansion in 3D or width in 2D. Fractures located at least partially
in one of these zones are labeled as entry or exit fractures, respectively.

\begin{figure}[H]
	\centering
	\includegraphics[width=0.5\textwidth]{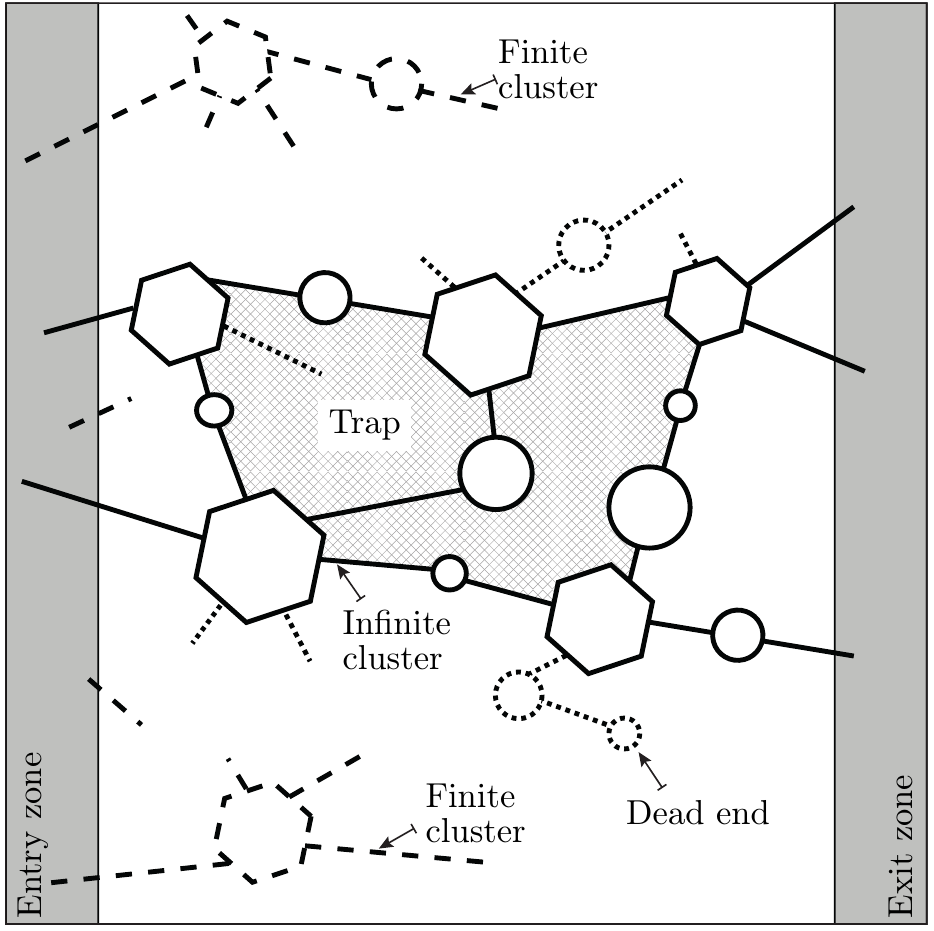}
	\caption[2D scheme of the breakthrough state]
		{Two dimensional scheme of the system at the breakthrough representing the infinite and some finite clusters.
		 The infinite cluster consists of a backbone (solid lines) and several dead
		 ends (dotted line). Finite clusters are represented by dashed lines and
		 trapped regions are hatched.}
	\label{fig: fig_infinite_cluster_scheme}
\end{figure}

After the forward percolation, the trapping MIP is applied, till the breakthrough step is
reached. At breakthrough, the largest interconnected cluster connects in- and outlet.
This breakthrough state is summarized schematically in Fig.~\ref{fig: fig_infinite_cluster_scheme}
with the finite clusters drawn by dashed lines and the infinite or percolation cluster by solid ones, connecting
the entry and exit zone marked gray. The backbone is drawn in
solid lines with links (chains of connected fractures) as lines, nodes (crossing points of the
links) as hexagons and blobs (dense regions with more than one connection between two points like
cycles or loops) as circles. The hatched area is representing the region containing trapped fractures.

Parts of the infinite cluster with dotted lines, connected to the cluster by only one bond,
resemble dead ends. Since they do not contribute to the actual fluid (e.g. gas) flow after
breakthrough, they can be pruned. Usually, the majority of the fractures form dead ends, but to
simplify the scheme, only very few of them are shown in Fig.~\ref{fig: fig_infinite_cluster_scheme}.
The finite clusters, which are not part of the
infinite cluster and will also not contribute to the flow process, are shown in dashed lines and can
be pruned as well. What remains is the so-called \textit{backbone}, being the portion of the infinite cluster
that actually contributes to the fluid flow. In studies of gas leakage and underground waste repositories,
the backbone is also called flow fracture network or leakage path. Some of the links are designated
\textit{red} or \textit{hot} bonds, since no alternate path exists and the spanning
cluster would become finite, if they were cut. All definitions are summarized in the
``links-nodes-blobs'' picture similar to Fig.~\ref{fig: fig_infinite_cluster_scheme}, introduced by Stanley \cite{Stanley1977}.

The backbone is identified using again a burning algorithm but starting from invaded
fractures at the exit zone and processing till it reaches invaded fractures in the entry zone.
Since this process seems to run backwards in terms of the spatial direction
given by the gas-water displacement simulated with MIP, it is called \textit{backward percolation}.
After this procedure, all fractures, that are not belonging to the identified infinite cluster are pruned.
Also fractures with only one connection, which are neither inlet nor outlet fractures, are deleted. This is done
recursively, till only the backbone remains.

After reaching the breakthrough state, there are several scenarios, which can be considered in addition.
Instead of stopping the MIP simulation at breakthrough, being followed by the backward percolation
as discussed above, the invading process can continue, i.e. by assuming an infinite gas production
rate at the inlet. The simulation can then be stopped either after \textit{full invasion} of the system,
meaning that no more water can be pressed out of the system and a specific water saturation
due to trapping has been reached, or after reaching a certain threshold value, e.g. representing the
highest possible input pressure at the inlet. Additional extensions could be pressure drops after breakthrough 
or above the percolation threshold after reaching again 
a certain threshold, to pretend a finite and maybe also dynamically varying gas production rate.
In this work, the full invasion scenario is investigated (Secs.~\ref{sec: sec_isotropic_anisotropic},
~\ref{sec: sec_comparison_wellenberg}), additional to the usual simulation studies till breakthrough.
All MIP variants or scenarios are followed by calculating post-breakthrough or
collapsed-network properties such as the invaded volume, gas-water saturation relationship, pressure distribution,
and flow rates.

%%%%%%%%%%%%%%%%%%%%%%%%%%%%%%%%%%%%%%%%%%%%%%%%%%%%%%%%%%%%%%%%%%%%%%%%%%%%%%%%%%%%%%%%%%%%%%%%
\section{Estimates for permeability after breakthrough}
	\label{sec: sec_permeability_breakthrough}
%%%%%%%%%%%%%%%%%%%%%%%%%%%%%%%%%%%%%%%%%%%%%%%%%%%%%%%%%%%%%%%%%%%%%%%%%%%%%%%%%%%%%%%%%%%%%%%%

An idealized fracture network can be regarded as a network of tiny connected tubes.
For small Reynolds numbers (of the order of one), \textit{Darcy's law} can be applied, which relates mass flow to pressure
gradient at the continuum scale. Darcy's law states, that the volumetric flow rate
$q$ of an incompressible fluid through a specimen of fractured porous material is given by
\begin{equation}
	q = \frac{k A \Delta p}{\mu l} \ ,
	\label{eq: darcy_law}
\end{equation}
where $\Delta p$ is the hydrostatic pressure difference across the specimen, $l$ is the
length of the specimen, $\mu$ the dynamic viscosity, and $A$ the cross sectional area. The constant of proportionality $k$
is called the \textit{Darcy permeability} of the material. Based on this simple
relationship, mass balance leads to an equation for pressure in the domain of interest.
The flow calculation through a single fracture using the parallel-plate concept from above
yields the \textit{cubic law}, namely
\begin{equation}
	q = \frac{w h^3 \Delta p}{12\mu l} \ ,
	\label{eq: cubic_law}
\end{equation}
where $q$ again denotes the volumetric flow rate, $\Delta p$ the pressure difference, $\mu$ the dynamic viscosity,
$w$ the width of the fracture, $l$ the distance between the inlet and outlet also called fracture length,
and $h$ the aperture of the fracture. The cubic law is a simplification of the Darcy's law in Eq.~(\ref{eq: darcy_law}).
The comparison of the cubic and Darcy's law shows that the permeability of the fracture can be identified as
$k = h^2/12$. This means that only the aperture is needed to calculate the permeability of a single fracture.

%----------------------------------------------------------------------------------------------%
\subsection{Visco-capillary two-phase flow in fractures}
	\label{subsec: subsec_visco-capillary_two_phase_flow}
%----------------------------------------------------------------------------------------------%

Visco-capillary two-phase flow occurs in a fracture network containing for example gas and water with a
meniscus separating the two phases. It is described as a transport process whereby the water in
the fracture or pore volume of a rock formation is displaced by gas under the influence of
viscous and capillary forces \cite{Bear1972}. Assuming again an idealized fracture network
regarded as a network of tiny connected capillary tubes, for each tube \textit{Young-Laplace's equation}
is used to describe the governing physics of a \textit{capillary pressure} $p_{\text{cap}}$ forming across the interface
between the two static fluids, gas and water, due to the surface tension
\begin{equation}
	p_{\text{cap}} = \displaystyle\frac{2\sigma\cos\theta}{h} \ .
	\label{eq: cappres}
\end{equation}
$\sigma$ denotes the surface or interfacial tension (for gas/water: $\sigma_{\text{gw}} \approx 0.0073$ Nm$^{-1}$),
$\theta$ is the wetting angle, and $h$ is the aperture value of the fracture. In other words,
Young-Laplace represents the difference between gas and water pressure needed to displace the water
from the initially fully saturated fracture network.
Relating the capillary pressure and the water saturation in the fracture network gives the
capillary pressure - water saturation relationship, commonly called \textit{water retention curve}.

%----------------------------------------------------------------------------------------------%
\subsection{Flow simulation in the channel model}
	\label{subsec: subsec_channel_model_discretization}
%----------------------------------------------------------------------------------------------%

Several approaches have been used to solve the flow equations stated above. Considering the case
of the DFN with a large number of fractures in 3D space, a simplification is inevitable.
Mostly, the description in the 3D network is replaced by a capillary model (tubes) as stated
before. One approach of a capillary model is called channel model and was proposed by
Cacas \textit{et al.} \cite{Cacas1990}. It holds the idea that the intersection of two fractures is schematized by a
channel (tube) that joins their centers with an effective hydraulic conductivity resulting from
simple geometric arguments. It is based on the observation that in real fracture planes, flow does
not occur over the entire surface of the fracture, but only in a limited number of channels within
the fracture plane. It further assumes that each intersection of the fracture planes
is connected to a fictitious node at the center of each circular fracture by a single equivalent
straight channel starting from the center of the intersection. The exact geometry of the real
channel is thereby omitted. In this work we use the centers of each intersection as fictitious nodes. The center
node is left out, since the flow in the fracture will always follow the direct path from node to node.

To solve the flow problem, a global system matrix for the flow fracture network, containing the elements
(channels) and nodes from the channel model is constructed. A similar equation for the cubic law in
Eq.~(\ref{eq: cubic_law}) can be expressed as
\begin{equation}
	\vec{Q} = \frac{1}{12\gamma} \textbf{M} \vec{P} \ ,
	\label{eq: cubic_law_system_matrix}
\end{equation}
where $\vec{Q}\in \mathbb{R}^m$ denotes the volumetric flow rate vector, $\vec{P}\in \mathbb{R}^n$ the pressure field vector,
$\gamma$ the dynamic viscosity and $\textbf{M}\in \mathbb{R}^{m\times n}$ the \textit{global system matrix}.
The permeability of the leakage path, seen as the permeability of an ``equivalent'' single fracture, can be identified as
\begin{equation}
	K_{\text{path}} = \frac{H_{\text{path}}^2}{12} \ ,
	\label{eq: tot_permeability}
\end{equation}
where $H_{\text{path}}$ is the equivalent or total aperture of the leakage path.
To summarize, results using the channel model are a very simple approximation of flow through geological
fracture networks. Nevertheless, the essential physics from the single fracture scale to the much larger fracture
network scale is included in the described model in a consistent and also computationally feasible way.
Note that the flow calculations are only performed on the backbones, that have to be found via the TMIP,
as shown in the following.

%%%%%%%%%%%%%%%%%%%%%%%%%%%%%%%%%%%%%%%%%%%%%%%%%%%%%%%%%%%%%%%%%%%%%%%%%%%%%%%%%%%%%%%%%%%%%%%%
\section{Isotropic vs. anisotropic fracture networks}
	\label{sec: sec_isotropic_anisotropic}
%%%%%%%%%%%%%%%%%%%%%%%%%%%%%%%%%%%%%%%%%%%%%%%%%%%%%%%%%%%%%%%%%%%%%%%%%%%%%%%%%%%%%%%%%%%%%%%%

\begin{figure}[H]
	\centering
	\includegraphics[width=1.0\textwidth]{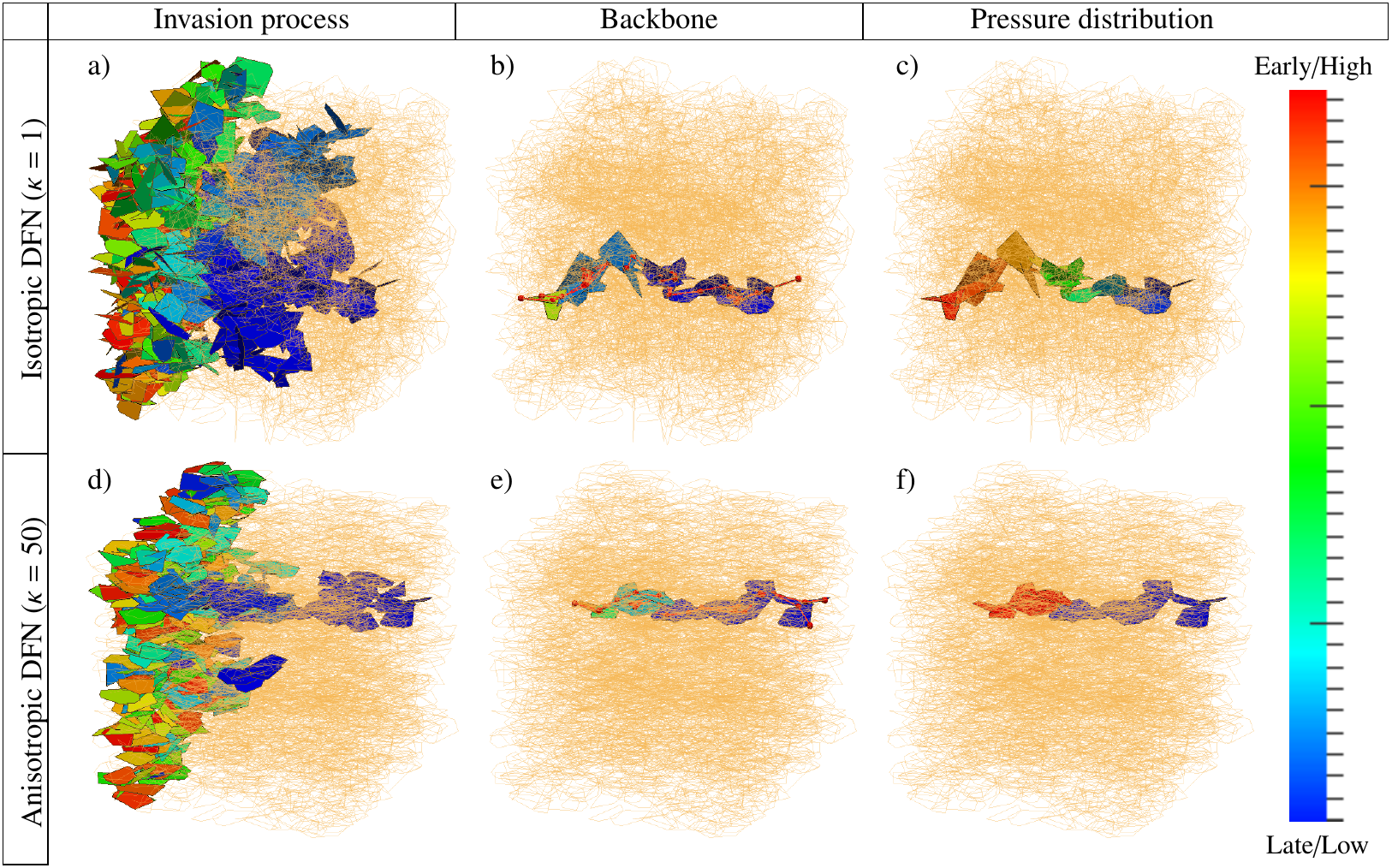}
	\caption[MIP on artificial 3D DFN - breakthrough]
		{DFN at \textbf{breakthrough}. \\
		 a) and d) Invaded fractures marked from early (red) to late invasion (blue)
		 after trapping MIP from left $y=0$ to right $y=L$.
		 b) and e) Flow fracture network (backbone of the percolation cluster) and
		 channel model discretization of the non-pruned fractures.
		 c) and f) Pressure distribution in the flow fracture network. The fractures are colored according
	   to the mean pressure on the fracture plane. Red indicates high and blue low pressure.
	   a), b), and c) correspond to an isotropic DFN ($\kappa=1$) while d), e), and f) are for an
	   anisotropic DFN ($\kappa=50$).}
	\label{fig: fig_TMIP_L10NP4000_breakthrough}
\end{figure}

\begin{figure}[H]
	\centering
	\includegraphics[width=1.0\textwidth]{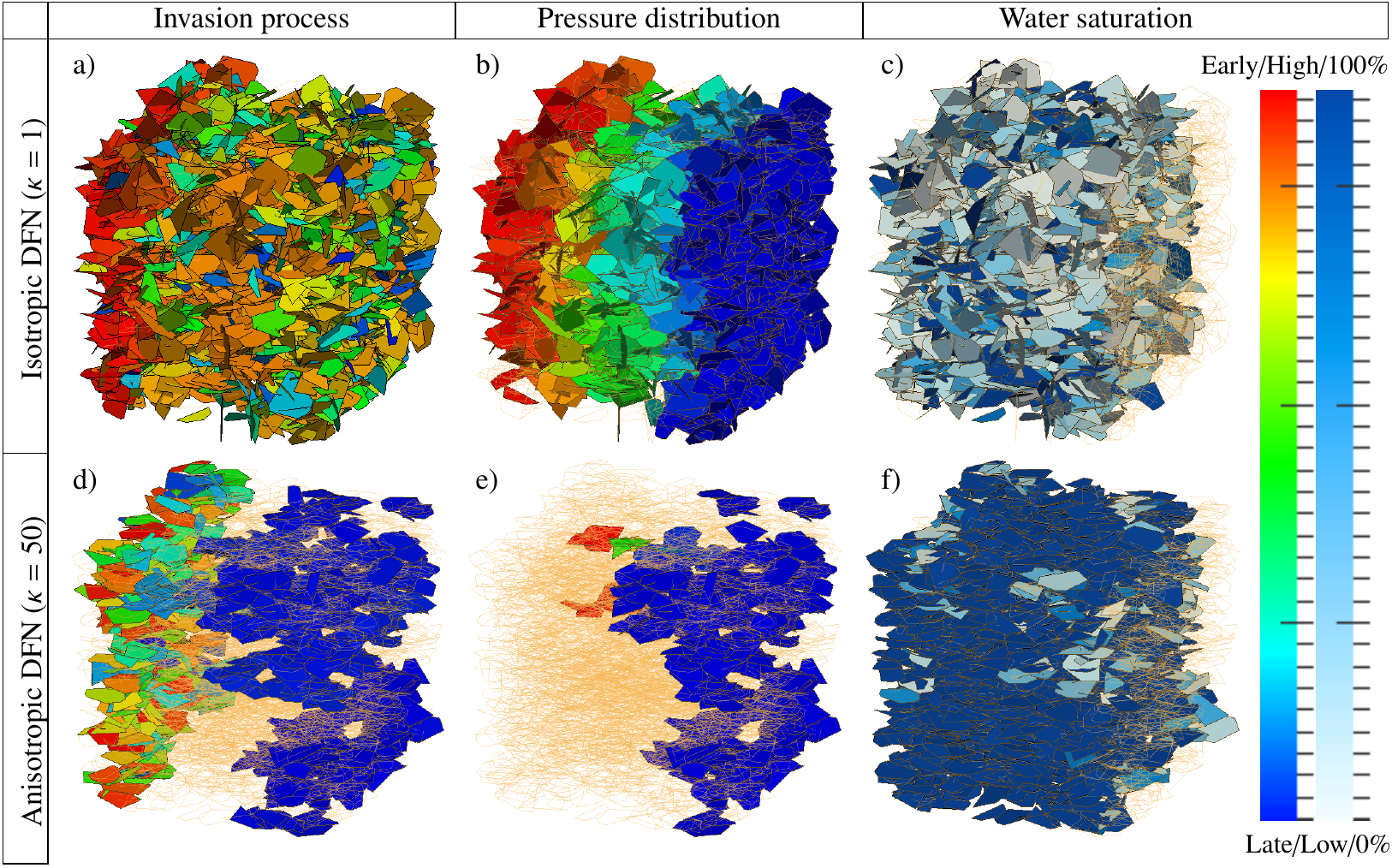}
	\caption[MIP on artificial 3D DFN - full invasion]
		{DFN after \textbf{full invasion} with trapping MIP. \\
		 a) and d) Invaded fractures.
		 b) and e) Pressure distribution in the flow fracture network.
		 c) and f) Invaded fractures that contain trapped water. Colors indicate the amount of trapped water
 		 from fully trapped with 100\% (dark blue) water content to non-trapped (fully invaded)
 		 with 0\% (white) water content. Transparent fractures at the exit zone (representing the outlet)
 	   are not identified (no trapping possible).}
	\label{fig: fig_TMIP_L10NP4000_full_invasion}
\end{figure}

To address the effect of anisotropy, trapping MIP ($C_{\text{I}}=1$) on the specific nearly
isotropic artificial 3D DFN (see Fig.~\ref{fig: fig_AFN_DFN_model} a)) and the highly anisotropic DFN
(see Fig.~\ref{fig: fig_AFN_DFN_model} b)) is studied. The DFNs are similar with respect to the
local fracture porosity and fracture area, only that the fracture orientation is changed by setting
the Fisher dispersion parameter from $\kappa=1$ for the isotropic case to $\kappa=50$ for the highly
anisotropic one. In all scenarios, the entry zone is defined as the $xz$-plane on the left side at $y=0$,
and the exit zone correspondingly as the $xz$-plane on the right side at $y=L$. Periodic boundaries
are applied on the other four sides. Accordingly, the invasion process starts at the entry zone and
progresses in the direction of the outlet. Instead of stopping the invasion process after reaching
the outlet, leading to the breakthrough in Fig.~\ref{fig: fig_TMIP_L10NP4000_breakthrough}, the process
continues by assuming an infinite gas production rate at the inlet. Finally, the process stops after
full invasion in Fig.~\ref{fig: fig_TMIP_L10NP4000_full_invasion}.

The concentration in terms of the average number of intersections per fracture is calculated numerically
as $\rho^{'}=\left\langle\overline{N_\text{int}}\right\rangle\approx 6.1$ for the isotropic ($\kappa=1$) and
$\rho^{'}\approx 1.93$ for the highly anisotropic ($\kappa=50$) DFN. Hence, the concentration of the anisotropic
DFN is actually considerably smaller than the one of the isotropic DFN. The
asymptotic values (for the infinite system size $L\rightarrow\infty$) found by Khamforoush \textit{et al.}
\cite{Khamforoush2008} for anisotropic DFNs with $\kappa=50$ are $\rho^{'}_{c^{\infty} x,A}\approx 2.1$
and $\rho^{'}_{c^{\infty} y,A}\approx 2.1$ in $x$- and $y$-direction and $\rho^{'}_{c^{\infty} z,A}\approx 2.4$
in $z$-direction. Given that the MIP proceeds in $y$-direction suggests a concentration
value slightly below the asymptotic value. Finite size effects are expected and are discussed
in Sec.~\ref{sec: sec_finite_size_dependence}.

The considerably smaller amount of fracture intersections (concentration) is immediately apparent comparing the
amount of invaded fractures after full invasion in Fig.~\ref{fig: fig_TMIP_L10NP4000_full_invasion} a) and d).
A large part, most notably in the middle of the DFN, is not invaded, leaving a very high proportion amount of trapped water in the DFN
(Fig.~\ref{fig: fig_TMIP_L10NP4000_full_invasion} c) and f)). The detailed description of the invasion process
is investigated in Sec.~\ref{sec: sec_comparison_wellenberg}, together with a real-world application.

%%%%%%%%%%%%%%%%%%%%%%%%%%%%%%%%%%%%%%%%%%%%%%%%%%%%%%%%%%%%%%%%%%%%%%%%%%%%%%%%%%%%%%%%%%%%%%%%
\section{Finite size dependence}
	\label{sec: sec_finite_size_dependence}
%%%%%%%%%%%%%%%%%%%%%%%%%%%%%%%%%%%%%%%%%%%%%%%%%%%%%%%%%%%%%%%%%%%%%%%%%%%%%%%%%%%%%%%%%%%%%%%%

\begin{figure}[H]
	\centering
	\includegraphics[width=0.6\textwidth]{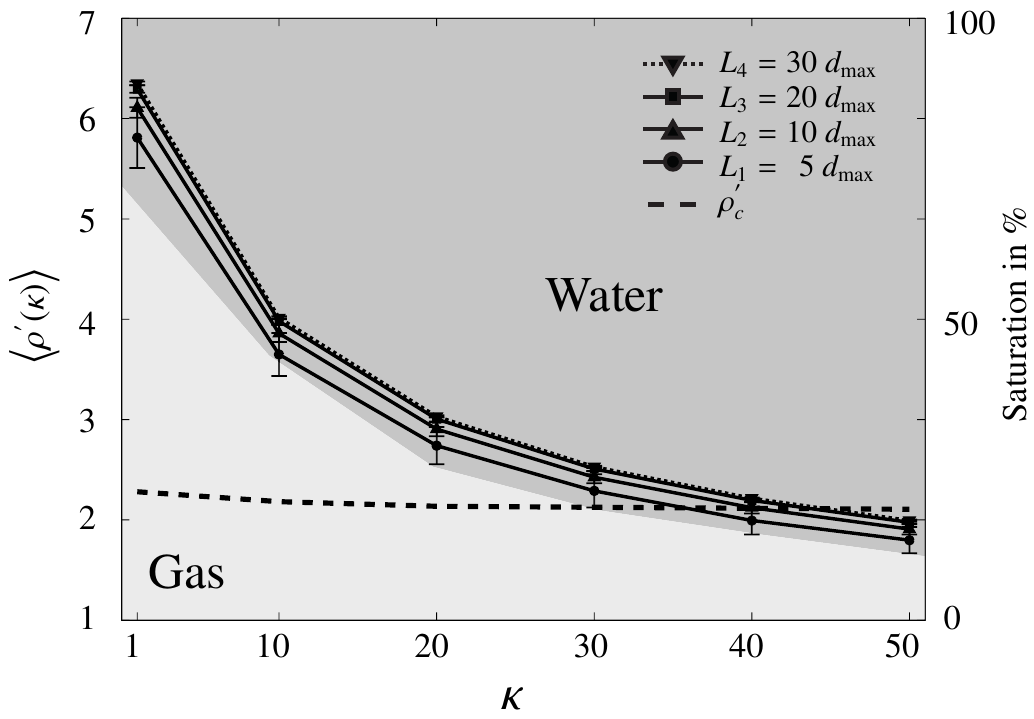}
  \caption[Average concentrations for DFN model system size]
		{Mean average concentration for DFNs with
		 varying system sizes, averaged over 1000 ($L_1$), 200 ($L_2$), and 100 ($L_3$ and $L_4$) realizations.}
	\label{fig: fig_TMIP_statplot_concentrations}
\end{figure}

To study size effects, trapping MIP is applied on several artificial 3D discrete fracture networks (DFN).
The DFNs are generated in a unit cell of size $L^3$. The polygons,
representing the fractures, are oriented using a varying Fisher dispersion parameter
$\kappa=\left\{1,10,20,30,40,50\right\}$. Four different system sizes are considered with system lengths
$L=\left\{5~d_\text{max},10~d_\text{max},20~d_\text{max},30~d_\text{max}\right\}$,
with the diameter of the circumscribed circle of the largest fracture in the DFN $d_\text{max}$.
To assure that the DFN models have a connectivity independent of $L$ and between entry and exit,
a fracture density of $\rho_{\text{fr}}=4$ was chosen, leading to DFN models slightly
above the percolation thresholds $p_c$ of ordinary percolation (OP) for the chosen system sizes
($L\leq\ 30~d_\text{max}$). The amount of fractures $N_\text{fr}$ for
$L=\left\{5~d_\text{max},10~d_\text{max},20~d_\text{max},30~d_\text{max}\right\}$ 
are therefore $N_\text{fr}=\left\{500,4000,32000,108000\right\}$ given that $N_\text{fr}=\rho_\text{fr}/L^3$.
However, it turns out that the concentration in terms of the average number of
intersections per fracture ($\rho^{'}=\left\langle\overline{N_\text{int}}\right\rangle$) is in the range of the
critical concentration $\rho^{'}_{c}$ for high $\kappa$ values. 
The mean average concentration $\left\langle\rho^{'}(\kappa)\right\rangle$
for different system sizes and varying $\kappa$ is calculated numerically and plotted
in Fig.~\ref{fig: fig_TMIP_statplot_concentrations}. Obviously, the resulting mean average
concentrations for $\kappa=40$ and $\kappa=50$ are below the critical concentration $\rho^{'}_c$
(illustrated by the the dashed line in Fig.~\ref{fig: fig_TMIP_statplot_concentrations}).
Additionally, the water saturation after full invasion with trapping MIP of DFNs with $L=10~d_\text{max}$
containing $4000$ fractures is considered. The gray coloring in Fig.~\ref{fig: fig_TMIP_statplot_concentrations}
shows a decreasing invaded or gas volume for increasing $\kappa$ value. The curve resembles the
decreasing concentration curve for $L=10~d_\text{max}$ satisfactorily.

\begin{figure}[H]
	\centering
	\includegraphics[width=1.0\textwidth]{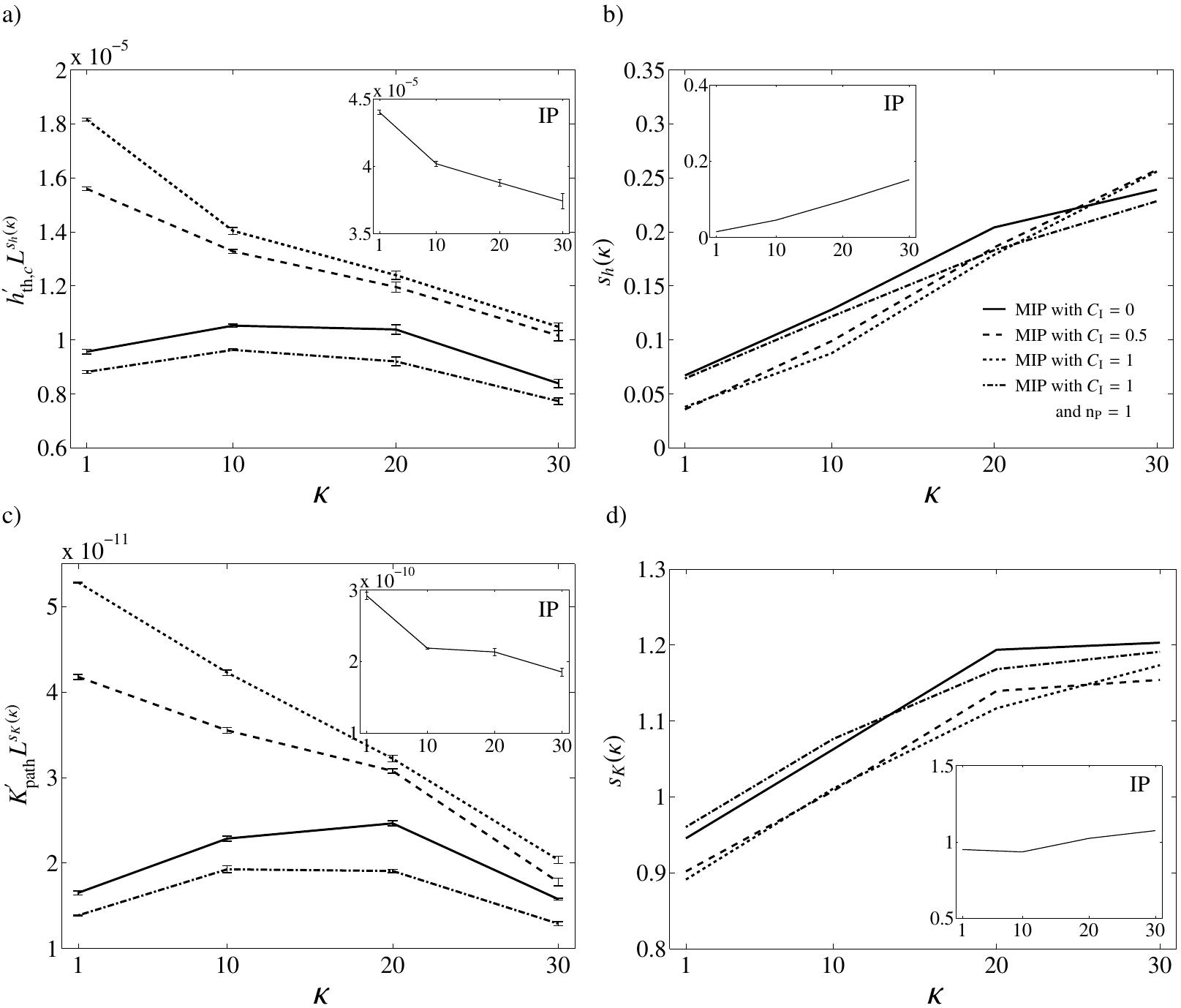}
  \caption[The finite size scaling of the aperture threshold and the permeability]
  	{a) The critical aperture threshold $h^{'}_{\text{th},c}$ rescaled by $L^{s_h(\kappa)}$ where $s_h(\kappa)$ is in b).
		 Trapping MIP with varying inclination adjustment constants $C_{\text{I}}$, one
		 with no path length adjustment ($\text{n}_{\text{P}}=1$), and ordinary trapping IP
		 with $\text{n}_{\text{P}}=\text{n}_{\text{C}}=\text{n}_{\text{I}}=1$ (small plot) are studied.
		 c) Similarly, the path permeability $K^{'}_{\text{path}}$ at breakthrough on the same DFN rescaled by
		 ${L^{s_K(\kappa)}}$ with $s_K(\kappa)$ in d).}
	\label{fig: fig_statplot_finite_size_dependence}
\end{figure}

To study the influence of the Fisher dispersion parameter $\kappa$ on the dimensionless critical aperture threshold
$h^{'}_{\text{th}}$ and on the dimensionless permeability of the flow path or backbone $K^{'}_{\text{path}}$
as well as to determine the scaling exponents of the different system sizes, the finite-size dependence is analyzed.
The fracture network permeability $K_{\text{n}}$ is expected to scale as \cite{Koudina1998}
\begin{equation}
	K_{\text{n}}(\rho^{'}_{c},L) \sim L^{-t/\nu} \ ,
	\label{eq: permeability_scaling_law}
\end{equation}
with conductivity exponent $t$ and correlation length exponent $\nu$. At the criticality, the exponents are believed
to be universal, i.e., to depend only on the dimension of the system, but not on the underlying lattice properties
\cite{Stauffer1994}. For DFNs, Koudina \textit{et al.} \cite{Koudina1998} report $t/v=2.22\pm0.08$, which is
in good agreement with the value for 3D $t/v\approx 2.26$ \cite{Normand1995}. Above the critical concentration, the exponent is no
longer universal and we found for $\kappa\leq 30$
\begin{equation}
	K^{'}_{\text{path}}(\kappa,L) \sim L^{-s_K(\kappa)} \ ,
	\label{eq: permeability_scaling_law_kappa}
\end{equation}
with dimensionless path permeability $K^{'}_{\text{path}}$ dependent on $\kappa$ and system length $L$. The dependency is given
by the nonlinear scaling function $s_K(\kappa)$. Similarly, a scaling law can be stated for the dimensionless critical
aperture threshold with
\begin{equation}
	h^{'}_{\text{th},c}(\kappa,L) \sim L^{-s_h(\kappa)} \ ,
	\label{eq: threshold_scaling_law_kappa}
\end{equation}
with the nonlinear scaling function $s_h(\kappa)$. The critical aperture threshold and the path permeability
for the artificial DFN rescaled accordingly are in Figs.~\ref{fig: fig_statplot_finite_size_dependence} a) and c),
and the functions $s_K(\kappa)$ and $s_h(\kappa)$ are in Figs.~\ref{fig: fig_statplot_finite_size_dependence} b) and d).
Trapping MIP without inclination adjustment shows a linear decrease for the path permeability
and a nonlinear decrease for the critical aperture threshold. This is in agreement with the results for
ordinary percolation (OP) found by Khamforoush \textit{et al.} \cite{Khamforoush2008}.

Trapping MIP with inclination adjustment shows an extremely nonlinear behavior. The reason for this is the strong
effect of the inclination adjustment for fractures, that are nearly perpendicular to each other for a small $\kappa$ value.
In other words, the inclination adjustment anticipates a higher critical aperture threshold and path permeability with
a decreasing strength for an increasing $\kappa$. This is the counterpart to reduction of the critical aperture threshold 
and path permeability with decreasing average concentration $\rho^{'}$ (increasing $\kappa$) shown 
in Fig.~\ref{fig: fig_TMIP_statplot_concentrations}. Both effects combined constitute this concave function with
a maximum between $\kappa=20$ and $\kappa=30$.

%%%%%%%%%%%%%%%%%%%%%%%%%%%%%%%%%%%%%%%%%%%%%%%%%%%%%%%%%%%%%%%%%%%%%%%%%%%%%%%%%%%%%%%%%%%%%%%%
\section{Comparison to ``real'' fracture network}
	\label{sec: sec_comparison_wellenberg}
%%%%%%%%%%%%%%%%%%%%%%%%%%%%%%%%%%%%%%%%%%%%%%%%%%%%%%%%%%%%%%%%%%%%%%%%%%%%%%%%%%%%%%%%%%%%%%%%

\begin{figure}[H]
	\centering
	\includegraphics[width=1.0\textwidth]{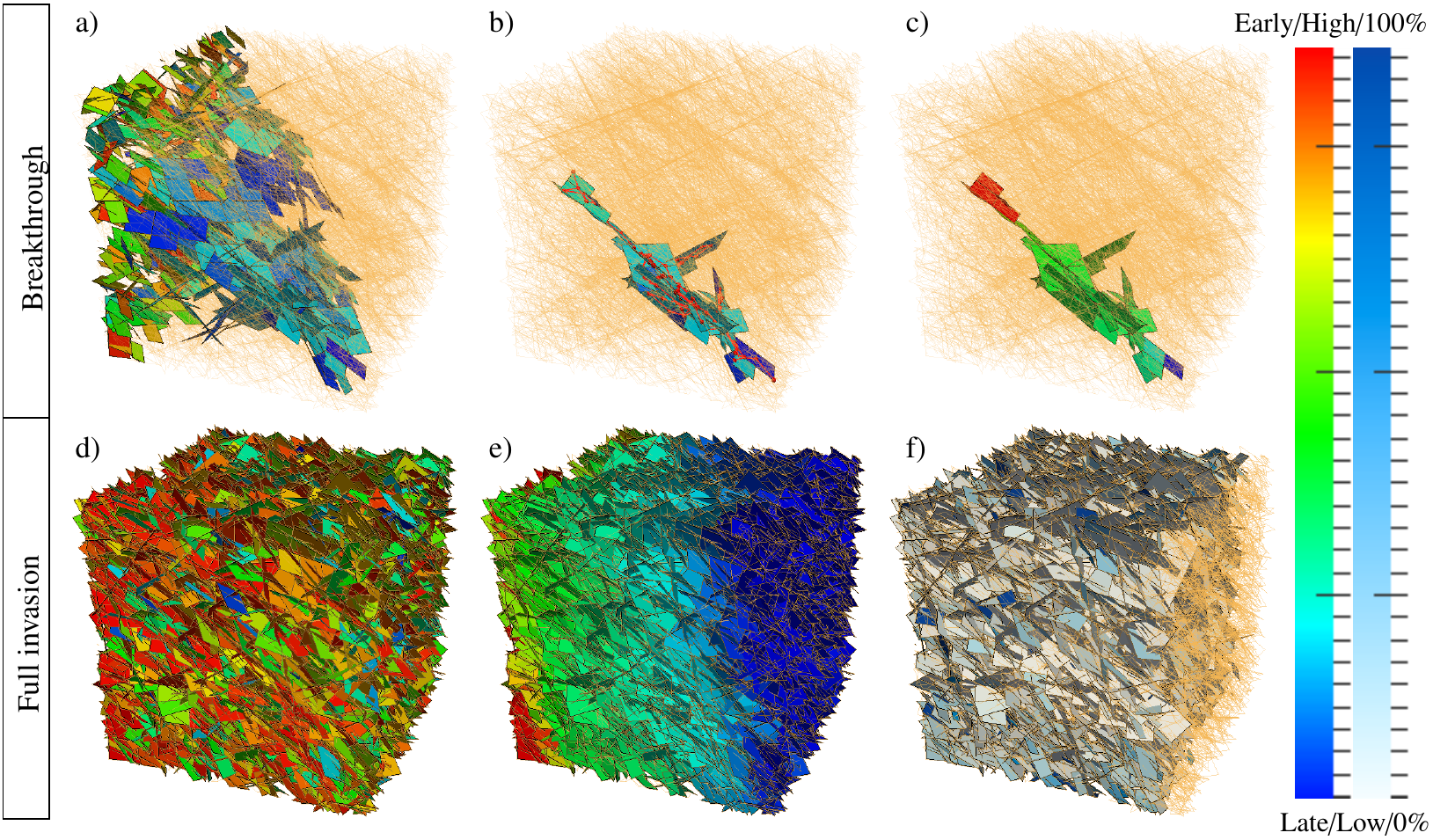}
	\caption[MIP on Wellenberg DFN]
		{Wellenberg DFN at \textbf{breakthrough} and after \textbf{full invasion}. \\
		 a) Invaded fractures at breakthrough marked from early (red) to late invasion (blue)
		 		after trapping MIP from left $y=0$ to right $y=100$~m.
		 b) Flow fracture network (backbone of the percolation cluster) and
		 		channel model discretization of the non-pruned fractures.
		 c) Pressure distribution in the flow fracture network. The fractures are colored according
	  		to the mean pressure on the fracture plane. Red indicates high and blue low pressure.
		 d) Invaded fractures after continuing the invading process till full invasion with trapping MIP.
		 e) Pressure distribution in the flow fracture network.
		 f) Invaded fractures that contain trapped water. Colors indicate the amount of trapped water
		 		from fully trapped with 100\% (dark blue) water content to non-trapped (fully invaded)
		 		with 0\% (white) water content. Transparent fractures at the exit zone (representing the outlet)
		 	  are not identified (no trapping possible).}
	\label{fig: fig_TMIP_wellenberg_invasion}
\end{figure}

The trapping MIP ($C_{\text{I}}=1$) was applied to the DFN model of the specific Wellenberg
realization with side length $L=100$~m containing $12684$ fractures. The invasion process started
at the left hand site at $y=0$~m and progressed to the right hand side at $y=100$~m. The breakthrough
and the full invasion state are shown in Fig.~\ref{fig: fig_TMIP_wellenberg_invasion}.
The full invasion state of this specific Wellenberg realization is studied and compared to the
artificially generated DFN models.

The decline of the aperture threshold for each invasion step can be used to study the dynamics of
the invading process. The decreasing aperture threshold illustrates thereby the rising entry pressure.
Additionally, the sum of invasions into fractures at each invasion step called \textit{invasion rate}
describes the advance of the gas migration in the DFN. Note that a single fracture might have several
inlets, represented by the connections to other fractures. Hence, more than one invasion into a single
fracture is possible and the total sum of invasion rates typically exceeds the total amount of fractures
for well connected DFNs. The aperture threshold value and the invasion rate for each step are shown
in Fig.~\ref{fig: fig_TMIP_wellenberg_invasion}. $d_\text{max}$ is set to $1$~m for the artificially
generated DFNs, to be able to directly compare the results to the Wellenberg DFN.

The breakthrough for the nearly isotropic DFN ($\kappa=1$) is reached after $918$ steps and at a critical aperture
(invasion percolation) threshold value of $h^{'}_{\text{th},c}\approx 7.41 10^{-6}$~m. The aperture
threshold curve, plotted in logarithmic scale, has a constant slope of $b\approx -1.27$ after breakthrough
till around $5000$ steps. For the highly anisotropic DFN ($\kappa=50$), breakthrough is reached after $553$ steps
and at a critical aperture (invasion percolation) threshold value of $h^{'}_{\text{th},c}\approx 4.77 10^{-7}$~m.
Hence the critical aperture threshold is more than one order of magnitude below the one of the isotropic case.
After breakthrough only $25$ more invasion steps are performed till the full invasion state is reached, where no more invasion
is possible. This is caused by the lack of connectivity of the anisotropic DFN and emphasizes,
that the system with infinite size is not percolating for $\rho^{'}_{\infty}<\rho^{'}_{c,\infty}$.

Apparently, a huge number of invasion steps have to be performed for the Wellenberg DFN till full invasion,
due to its high connectivity. The invasion rate is defined as the sum of all invasions into fractures.
The faults in the Wellenberg DFN model are tessellated into equally sized fractures. Hence, multiple invasions
into fractures belonging to the same macro fracture (same fault) can occur at the same step.
The natural behavior of taking the path of the least resistance might lead to preferred spreading in macro
fractures for a certain amount of steps. A closer inspection of Fig.~\ref{fig: fig_TMIP_wellenberg_invasion}
a) and b) with respect to the invasion step coloring of macro fractures supports this assumption. The resulting
capillary pressure is calculated using Eq.~(\ref{eq: cappres}) and plotted in relation to the water saturation.
As assumed, only a small amount of trapped water remains in the Wellenberg DFN. The aperture threshold and water retention
curves of the isotropic and the Wellenberg DFN are very similar, with slightly steeper constant slope ($b\approx 1.71$)
between breakthrough and around $45000$ steps for the Wellenberg DFN. Although the Wellenberg DFN is of anisotropic nature.
Also the water retention curve (capillary pressure - water saturation curve) resembles the one of the nearly isotropic
artificial DFN in both shape and values. The only difference is that the remaining water saturation of the Wellenberg DFN
at full invasion is about $10$\% smaller than in the artificial one. This confirms the above-mentioned assumption, that
the higher connectivity of the DFN leads to a smaller amount of trapped water after full invasion.

\begin{figure}[H]
	\centering
	\includegraphics[width=1.0\textwidth]{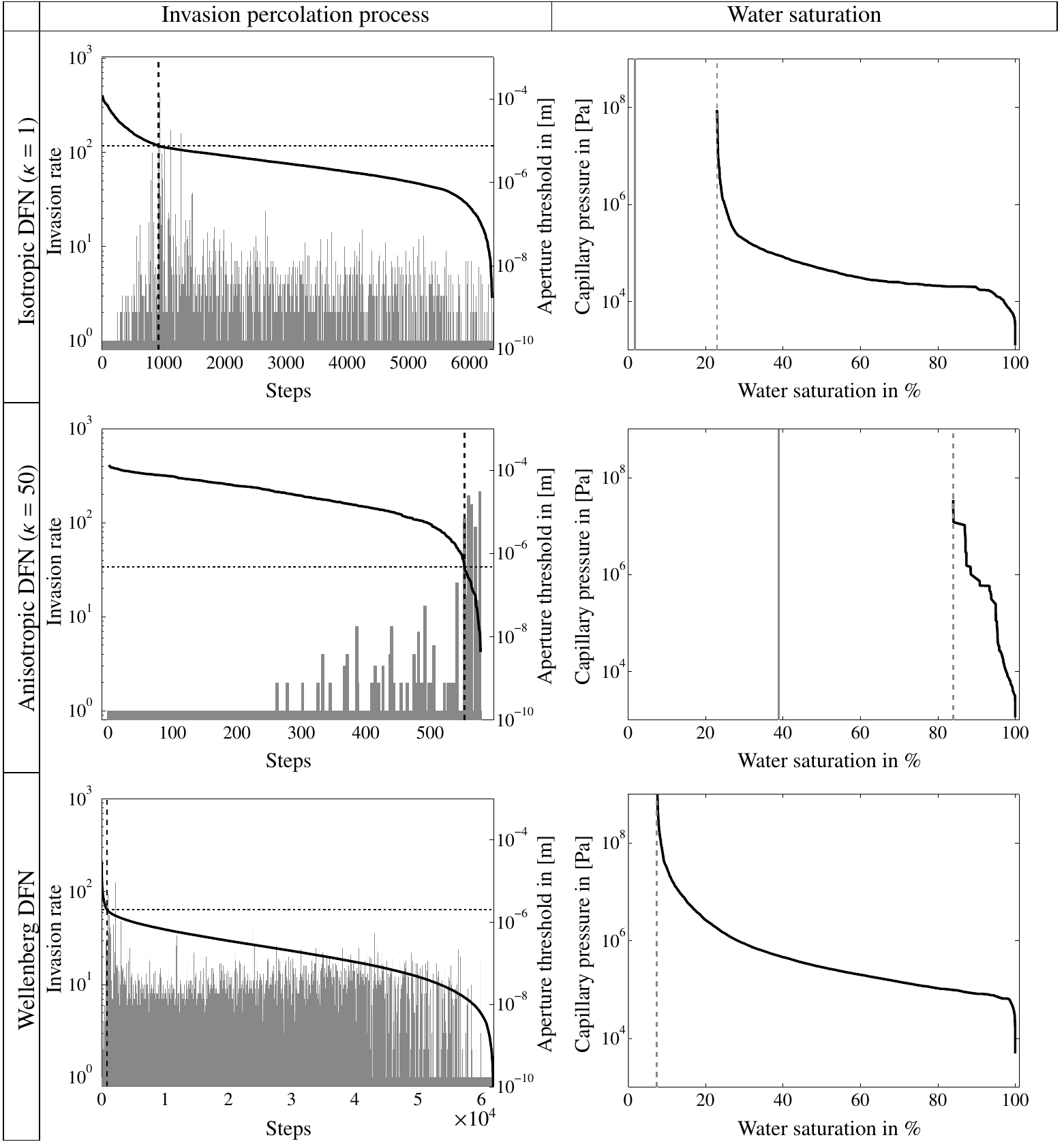}
  \caption[Comparison artificial and Wellenberg DFN]
		{On the left: The invasion process with the aperture threshold and the invasion rate for each step for trapping MIP on the
		 three different DFNs. The breakthrough step is marked by a dashed line for the invasion steps and by
		 a dotted line for the aperture threshold value.
		 On the right: Capillary pressure - water saturation relationship with the total amount of trapped water
		 marked by a dashed line. The amount of trapped water of pruned ``dead end'' fractures, that are not invadable
		 (determined before the MIP) is indicated by a solid cross line (perpendicular to the saturation axis).}
	\label{fig: fig_comp_ADFN_Wellenberg}
\end{figure}

%%%%%%%%%%%%%%%%%%%%%%%%%%%%%%%%%%%%%%%%%%%%%%%%%%%%%%%%%%%%%%%%%%%%%%%%%%%%%%%%%%%%%%%%%%%%%%%%
\section{Conclusions}
	\label{sec: sec_conclusions}
%%%%%%%%%%%%%%%%%%%%%%%%%%%%%%%%%%%%%%%%%%%%%%%%%%%%%%%%%%%%%%%%%%%%%%%%%%%%%%%%%%%%%%%%%%%%%%%%

This work has investigated two-phase flow in anisotropic three-dimensional (3D) discrete fracture
networks (DFN). The DFNs are either based on a general distribution function or have been conditioned
on geological measurements from boreholes, core and outcrop. A modified invasion percolation (MIP)
was proposed to study percolation of the immiscible fluids through the DFN. The MIP was designed to
describe the behavior at fracture intersections in more detail by including the contact angles and the
length of the line of intersection between connected fractures. Furthermore, it incorporates the hydraulic
path length from one connection point to the other.

A systematic study of the critical aperture threshold and the path permeability of the remaining
flow fracture network after trapping MIP on artificial anisotropic 3D DFNs is shown in this work.
The DFNs are constructed according to the Fisher distribution. The anisotropy is thereby controlled
by the Fisher dispersion parameter $\kappa$. With increasing $\kappa$, the distribution of the
fracture normal vector becomes more dense and inclined towards the initial mean direction, chosen to be
the $z$-axis. It is found for trapping MIP without inclination adjustment, that increased anisotropy
in $z$-direction leads to a decreased critical aperture (invasion percolation) threshold.
The dependency of the spanning path (infinite cluster) to the anisotropy
is in good agreement with the study of Hunt \textit{et al.} \cite{Hunt2006} and Khamforoush \textit{et al.}
\cite{Khamforoush2008}. Furthermore, the path permeability of the remaining flow fracture networks is
decreased similarly with increased anisotropy. Including the inclination adjustment anticipates a
higher critical aperture threshold and path permeability with a decreasing strength for increasing
$\kappa$. This is the counterpart to the reduction of the critical aperture threshold and path permeability
with increasing $\kappa$. Both effects combined constitute a concave function, that increases with $\kappa$
until a certain value depending on the average concentration in terms of the average number of intersections
per fracture and decreases afterwards. Additional modifications of the invasion percolation procedure,
like gravity adjustment, might have a similar influence on the critical aperture threshold and path
permeability to $\kappa$ curves. These findings are in good agreement with our physical perception
and can be used for extremely anisotropic DFNs.

The ``real'' application, the so-called Wellenberg DFN model, was found to be directly comparable
to the artificial 3D DFN. Due to the high connectivity of the Wellenberg DFN model, it resembles mostly
the nearly isotropic artificial DFN. The main difference is found to be the remaining water saturation
of the Wellenberg DFN at full invasion, which is smaller than in the artificial one. This is due to
the higher connectivity of the DFN that avoids trapping during the MIP process. It was shown in Ref.~\cite{Wilkinson1983},
that trapping in general 3D lattices is very rare. Trapping MIP on fully isotropic DFN with extremely high
connectivity is expected to show the same characteristics. This was demonstrated for many applications of
porous media \cite{Sahimi1994,Stauffer1994,Hunt2005}. Note that the isolated boundaries of the Wellenberg
model in comparison to the periodic boundaries of the artificial DFNs are unproblematic due to the
high connectivity of the Wellenberg.

It is our belief, that invasion percolation with physical modifications is a useful numerical tool
for the physical abstraction of the complicated situation of two-phase flow in fractured rock. It helps
to reduce the problem to flow in a percolation backbone containing a very small portion of the total amount of fractures,
that could potentially be invaded. Further studies should focus on the comparison with more detailed physics-based models,
volumetric calculations with fluid dynamics of flow in typical situations like fracture intersections with measured geometries
to advance the inclination adjustments, or two-phase flow inside real fracture cavities in specific
rock as such. Since percolation features are strongly affected by correlations in the media
\cite{Herrmann1993,Sahimi1994b,Sahimi1996,Makse1996,Araujo2002,Araujo2003,Oliveira2011},
it would also be relevant to analyze the impact of spatially correlated distributions of apertures
and fracture orientations on the flow properties in successive work.

%%%%%%%%%%%%%%%%%%%%%%%%%%%%%%%%%%%%%%%%%%%%%%%%%%%%%%%%%%%%%%%%%%%%%%%%%%%%%%%%%%%%%%%%%%%%%%%%
\section*{Acknowledgments}
	\label{sec: sec_acknowledgments}
%%%%%%%%%%%%%%%%%%%%%%%%%%%%%%%%%%%%%%%%%%%%%%%%%%%%%%%%%%%%%%%%%%%%%%%%%%%%%%%%%%%%%%%%%%%%%%%%

The authors are thankful to Paul Marschall (Nagra), for fruitful discussions and ongoing support.

%~~~~~~~~~~~~~~~~~~~~~~~~~~~~~~~~~~~~~~~~~~~~~~~~~~~~~~~~~~~~~~~~~~~~~~~~~~~~~~~~~~~~~~~~~~~~~~%

%~~~~~~~~~~~~~~~~~~~~~~~~~~~~~~~~~~~~~~~~~~~~~~~~~~~~~~~~~~~~~~~~~~~~~~~~~~~~~~~~~~~~~~~~~~~~~~%
% bibliography, end document
%~~~~~~~~~~~~~~~~~~~~~~~~~~~~~~~~~~~~~~~~~~~~~~~~~~~~~~~~~~~~~~~~~~~~~~~~~~~~~~~~~~~~~~~~~~~~~~%
\biboptions{sort&compress}
\bibliographystyle{model1a-num-names}
\bibliography{literature}
\end{document}